\documentclass[aps,prb,twocolumn,superscriptaddress]{revtex4-2}
\usepackage{amssymb, amsmath, bm}
\usepackage{graphicx,onlyamsmath}
\usepackage[utf8]{inputenc}

\usepackage{siunitx}
\usepackage[version=4]{mhchem}

\usepackage{natbib}
\usepackage{xcolor}
\usepackage[normalem]{ulem}  
\UseRawInputEncoding

\usepackage{mathtools}
\usepackage{graphicx}
\usepackage{mathrsfs}
\usepackage{amssymb}
\usepackage{xcolor}
\usepackage{amsmath}
\usepackage{bm}
\usepackage{amsmath}
\usepackage{physics}
\makeatletter\AtBeginDocument{\let\@elt\relax}\makeatother
\usepackage{amsmath}
\usepackage[normalem]{ulem}
\usepackage[unicode=true,colorlinks=true,citecolor=blue,urlcolor=blue]{hyperref}
\usepackage{makecell}

\def\be {\begin{equation}}
	\def\ee {\end{equation}}
\def\bea {\begin{align}}
	\def\eea {\end{align}}

\def\bee{\begin{eqnarray}}
	\def\eee{\end{eqnarray}}
\def\BC {\begin{cases}}
	\def\EC {\end{cases}}

\newcommand{\re}[1]{\textcolor{red}{#1}}
\newcommand{\LG}[1]{\textcolor{magenta}{#1}}

\makeatother

\begin{document}

\title{Positive Terahertz Photoconductivity in  CdHgTe Under Hydrostatic Pressure\\
}

\author{I. Yahniuk}
\affiliation{Terahertz Center, University of Regensburg, D-93053 Regensburg, Germany}
	\affiliation{CENTERA Labs, Institute of High Pressure Physics, PAS, 01 - 142 Warsaw, Poland}
	
	\author{D.A. Kozlov}
	\affiliation{Terahertz Center, University of Regensburg, D-93053 Regensburg, Germany}
	
\author{M.D. Moldavskaya}
\affiliation{Terahertz Center, University of Regensburg, D-93053 Regensburg, Germany}

\author{L.E. Golub}
\affiliation{Terahertz Center, University of Regensburg, D-93053 Regensburg, Germany}

\author{V.V. Bel'kov}
\affiliation{Terahertz Center, University of Regensburg, D-93053 Regensburg, Germany}

\author{I.A. Dmitriev}
\affiliation{Terahertz Center, University of Regensburg, D-93053 Regensburg, Germany}

\author{S.S.~Krishtopenko}
\affiliation{Laboratoire Charles Coulomb (L2C), UMR 5221 CNRS-Universit\'{e} de Montpellier, F-34095 Montpellier, France}

\author{F. Teppe}
\affiliation{Laboratoire Charles Coulomb (L2C), UMR 5221 CNRS-Universit\'{e} de Montpellier, F-34095 Montpellier, France}

\author{Y.~Ivonyak}
\affiliation{CENTERA Labs, Institute of High Pressure Physics, PAS, 01 - 142 Warsaw, Poland}
\affiliation{CENTERA2, CEZAMAT, Warsaw University of Technology, 02 - 822 Warsaw, Poland}

\author{A.~Bercha}
\affiliation{CENTERA Labs, Institute of High Pressure Physics, PAS, 01 - 142 Warsaw, Poland}

\author{G.~Cywi\'{n}ski}
\affiliation{CENTERA Labs, Institute of High Pressure Physics, PAS, 01 - 142 Warsaw, Poland}
\affiliation{CENTERA2, CEZAMAT, Warsaw University of Technology, 02 - 822 Warsaw, Poland}

\author{W.~Knap}
\affiliation{CENTERA Labs, Institute of High Pressure Physics, PAS, 01 - 142 Warsaw, Poland}
\affiliation{CENTERA2, CEZAMAT, Warsaw University of Technology, 02 - 822 Warsaw, Poland}
	
\author{S.D.~Ganichev}
\affiliation{Terahertz Center, University of Regensburg, D-93053 Regensburg, Germany}
	\affiliation{CENTERA Labs, Institute of High Pressure Physics, PAS, 01 - 142 Warsaw, Poland}

%
%
%
%
%
%
%


\begin{abstract}
Positive terahertz photoconductivity is observed at room temperature in CdHgTe thin films with different Cd contents. We show that electron gas heating caused by Drude-like absorption results in positive photoconductivity because of the interband activation mechanism specific for undoped narrow-gap semiconductors and semimetals. 
Applying intense terahertz radiation, we observed that the photoconductivity saturates at high intensities, which was found to be caused by absorption bleaching. 
Both the magnitude of the photoconductivity and the saturation intensity are shown to exhibit an exponential dependence on the hydrostatic pressure. We show that this is a consequence of the fact that both phenomena are controlled by the ratio of energy and momentum relaxation times.
\end{abstract}

\maketitle

\section{\label{sec:Int}Introduction}

For several decades, bulk Cd$_x$Hg$_{1-x}$Te, often called MCT, has been considered as one of the best materials for infrared photodetectors~\cite{Capper1997,Norton2002,Henini2002,Rogalski2005,Downs2013,Rogalski2018,Vanamala2019}. In addition, it is known as a good material for the detection of terahertz (far-infrared) radiation~\cite{Dvoretsky2010,Rumyantsev2017,Ruffenach2017,Yavorskiy2018,Bak2018}. Although these materials have been extensively studied, they are still the focus of current research. The reason for this, besides the application point of view, is that CdHgTe alloys host new elementary quasiparticles named Kane fermions that demonstrate unique pseudo-relativistic behavior~\cite{Orlita2014}. Their bandgap, and hence their rest mass, is tunable as a function of temperature, chemical composition, pressure, or disorder, while their velocity remains unaffected~\cite{Teppe2016,Laurenti1990,Harman1961,Szola2022,Krishtopenko2022}. In the inverted phase, Kane fermions with a negative rest mass coexist with 2D Volkov-Pankratov states at the material interfaces~\cite{Pankratov1987,Krishtopenko2020}, which have also been observed and characterized in Refs.~\cite{Kazakov2021,Savchenko2023,Inhofer2017,Otteneder2020a}, further enhancing their remarkable characteristics. Unlike typical Dirac fermions, the energy dispersion of Kane fermions is characterized by Dirac cones with an additional flat band at the vertex. Recent research suggests that these quasiparticles can be viewed as ``complex'', composed of two nested Dirac states with different rest masses, highlighting their non-trivial nature \cite{Krishtopenko2022a}. Kane fermions also exhibit specific transport properties, with mobilities reaching up to $10^6$~cm$^2$/(V$\cdot$s) near the topological phase transition~\cite{Yavorskiy2018} and a giant nonsaturating magnetoresistance resulting from the gap opening in their Landau dispersion~\cite{Vasileva2020}. They also exhibit strong nonlinear THz dynamics, which has been recently observed~\cite{Soranzio2024}. Moreover, their relativistic properties also suppress nonradiative Auger recombination between their nonequidistant Landau levels, suggesting a promising application for THz Landau lasers~\cite{But2019}. All this gives rise to a new field of activities in the study of these crystals~\cite{Galeeva2017,Galeeva2020,Hubmann2020,Dvoretsky2019,Varavin2020}. Studies of radiation-induced changes in material conductivity, which is the basic principle of infrared MCT detectors, provide important insights into material properties.

\begin{figure}[t]
	\includegraphics [width=\columnwidth, keepaspectratio] {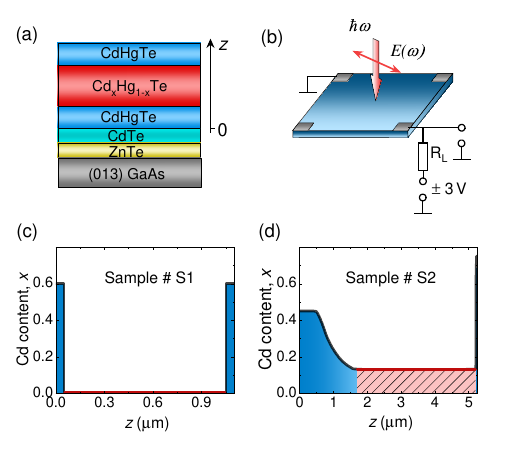} 
	\caption{\label{Fig:1} (a) A cross-sectional sketch of the investigated  Cd$_x$Hg$_{1-x}$Te structures. (b) The measurement configuration. (c) and (d) The distribution of the cadmium content $x$ in the near-surface layers of studied  samples as a function of coordinate $z$ measuring the distance towards the surface from the bottom CdTe layer, see panel (a).  Red lines and shaded area highlight the active layers of thickness $d$ with constant Cd content: for sample \#S1 $x=0$ and $d=1.0~\mu$m, for sample \#S2 $x=0.13$ and $d=3.45~\mu$m. 		The thickness of the GaAs substrate is 600~$\mu$m, the ZnTe adhesion layer is 30~nm, and the CdTe buffer layer is 5~$\mu$m. 
	}
\end{figure}

Here we report on the observation and study of THz radiation-induced photoconductivity at room temperature in Cd$_x$Hg$_{1-x}$Te alloys, exploring the effect of hydrostatic pressure on samples with different concentrations of Cd. We show that, surprisingly, THz radiation leads to positive photoconductivity, reflecting a decrease in resistance. It is also shown that the photoconductivity signal increases exponentially, growing by more than an order of magnitude at high pressures (up to 18 kbar). By studying the photoconductive response for different radiation intensities $I$, we also observed that the photoconductivity saturates as $I$ increases.  We found that the saturation intensity decreases exponentially with the pressure increase. Importantly, the exponents for photoconductivity magnitude and saturation intensities have the same magnitude but opposite sign. The results are obtained for the alloys with $x=0$ and $x=0.13$, which are characterized by qualitatively different band structures.  Our analysis shows that the observed positive photoconductivity and its saturation are caused by the electron gas heating, which increases the density of the free carriers contributing to the radiation absorption. The analysis demonstrates that as the pressure goes up, there is an exponential increase in $\Delta T = T_e -T_0 \propto \tau_\varepsilon/\tau_p$. Here $T_e$ and $T_0$ are the electron and the lattice temperatures, $\tau_\varepsilon$ and $\tau_p$ are the energy and momentum relaxation times. The obtained results are useful for the development of MCT-based THz detectors and defining their dynamic range.

\begin{figure}[t]
		\includegraphics [width=\columnwidth, keepaspectratio] {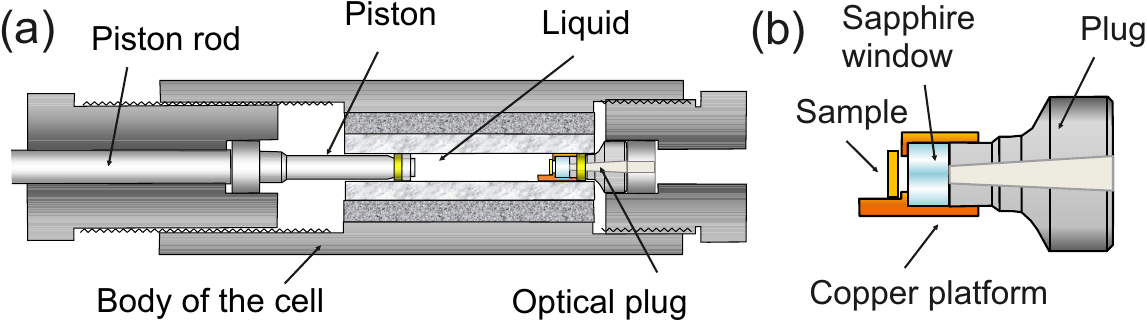} 
	\caption{\label{Fig:1pressure}  (a) Sketch of the pressure cell. (b) A zoomed image of the optical plug with the sample end window.}
\end{figure}

\section{\label{sec:Exp} Samples and methods}

\subsection{Samples and pressure cell}

\begin{figure}[t]
	\includegraphics [width=\columnwidth, keepaspectratio] {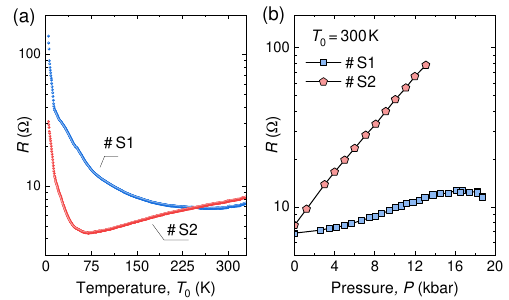}
	\caption{\label{Fig:2res} Temperature (a) and pressure (b) dependences of the two-terminal resistance $R$. 
				}
	\label{Fig_R_vs_T_P}
\end{figure}

The photoconductivity and its dependence on the hydrostatic pressure have been studied in  Cd$_x$Hg$_{1 - x}$Te films molecular beam epitaxy (MBE) grown on semi-insulating (013)-oriented GaAs substrates. The layer sequence and the band profile as a function of the distance to the film surface are shown in Fig.~\ref{Fig:1}. Two samples  with different values of the Cd contents $x$ were used: $x=0$ (sample \#S1), and $x=0.13$ (sample \#S2). All experiments were performed at room temperature, where sample~\#S1 exhibited a semimetallic behavior with an inverted band structure and zero forbidden gap, while sample~\#S2 had a non-inverted band structure. In sample \#S2, the band gap varied from about 50  to 140 meV as the pressure $P$ is increased from 1 bar to 11 kbar. The active layers with constant $x$ contents (red shaded area in Fig.~\ref{Fig:1}) were surrounded by regions with increasing cadmium concentrations. A CdTe buffer layer was needed to obtain the desired crystal quality for subsequent growth of CdHgTe. Due to thermal activation at room temperature, the samples contained bulk electrons and holes with concentrations ranging from $10^{16}$ to $3\times 10^{17}$\,cm$^{-3}$, for more detail see Sec.~\ref{discussion}.

All wafers were cut into square samples with sizes of $5\times5$~mm$^2$ or $2\times2$~mm$^2$ for ambient and high pressure experiments, respectively. Four ohmic indium contacts have been soldered to the sample corners, see Fig.~\ref{Fig:1}(b). Figure~\ref{Fig:2res}(a) shows the temperature dependences of the  two-point resistance $R$ measured at atmospheric pressure.  In the temperature range from 4 to 30~K, the resistance of both samples exhibits similar behavior, with a sharp decrease by an order of magnitude. Upon temperature increase, sample \#S1 shows a gradual decrease in resistance with a broad minimum in the range from 225 to 300~K. By contrast, sample \#S2 displays a more pronounced minimum at 60~K, followed by a clear increase. The shape of the curves is determined by the opposing effects of temperature on carrier density and mobility, which differ slightly between samples \#S1 and \#S2 and will be discussed in Sec.~\ref{discussion}. The resistivity value measured using the 4-point scheme on samples similar to those studied here are fully reproducible with the temperature dependence in Fig.~\ref{Fig_R_vs_T_P} of our work after deducting the constant wire resistance equal to 0.45~Ohm and considering the geometric factor equal to 2. This confirms the insignificance of the contribution of contact resistances in our case.

More details on the sample's composition, parameters and low temperature transport properties can be found in Refs.~\cite{Otteneder2020a, Moldavskaya2024, Savchenko2023, Savchenko2024} where the samples made from the same batches were studied.

\begin{figure}[t]
	\includegraphics [width=\columnwidth, keepaspectratio] {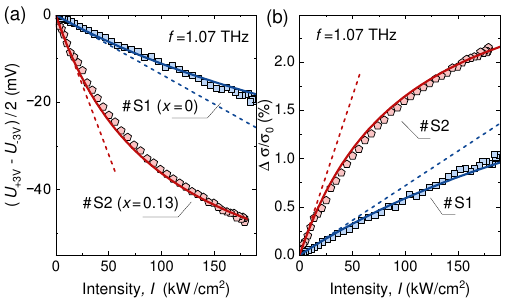}
	\caption{\label{Fig:2}  Intensity dependence of (a) the photoconductive signal $U=(U_{+3V}-U_{-3V})/2$ and (b) the relative change of the conductivity $\Delta\sigma/\sigma_0$. 	Solid lines represent fits after Eqs.~\eqref{KDr} and ~\eqref{DSSNL},  using $U \propto \Delta\sigma/\sigma_0 \propto K(I)\times I$. Dashed lines in panel (b) show the linear fit matching the low-intensity part of Eq.~\eqref{DSSNL}, $\Delta\sigma/\sigma_0 = A_{\rm 0} \times I$, and in panel (a) $U \propto I$.} 
\end{figure}

For experiments under hydrostatic pressure $P$, the small size samples were mounted in a specially designed pressure cell. A schematic illustration of the pressure equipment used for experiments is shown in Fig.~\ref{Fig:1pressure}. The sample was placed on the holder, which was then mounted on a copper heat sink and an optical pressure plug, see Fig.~\ref{Fig:1pressure}(a). Investigated samples were fixed at a distance of a few millimeters in front of the sapphire window.  The THz beam passed through the 1.2 mm diameter window and through the compressed liquid.  The pressure medium used was transformer oil and kerosene (in a 1:1 ratio), which remains semi-transparent in the THz frequency range under hydrostatic pressure. The design of the metallic pressure cell, see Fig.~\ref{Fig:1pressure}(b), allowed us to reach a pressure of up to 20 kbar. For slow pressure tuning and control, the cell was placed at a specific press with a force generator, which allowed us to change the force on the piston gradually.  Figure~\ref{Fig:2res}(b) shows the dependence of the two-point resistance on the hydrostatic pressure $P$. The observed increase in resistance is associated with the modification of the band structure under hydrostatic pressure and will be discussed in detail in Sec.~\ref{discussion}.

\begin{figure}[t]
	\includegraphics [width=\columnwidth, keepaspectratio] {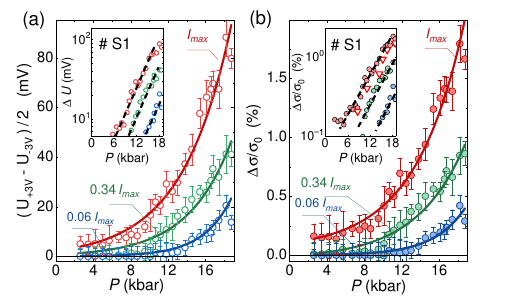}
	\caption{Pressure dependences of (a) the photoconductive signal $U = (U_{\rm +3V} -  U_{\rm -3V})/2$ and (b) the normalized photoconductivity $\Delta \sigma / \sigma_0$. The data were obtained for sample \#S1 and different values of the radiation intensity, indicated by the color code:  $I_{\rm max}=3.2$~kW/cm$^2$ (red), $0.34\,I_{\rm max}$ (green), and $0.06\,I_{\rm max}$ (blue). The insets show the results on a logarithmic scale. In addition to the data of the main panel we present the data from Fig.~\ref{Fig:4s1}(a) shown by down red triangles obtained for $I=3$~kW/cm$^2$. The solid lines show data fits with an exponential function.}
	\label{fig_PC_vs_P_normalizedS1}
\end{figure}

\begin{figure}[t]
\includegraphics [width=\columnwidth, keepaspectratio] {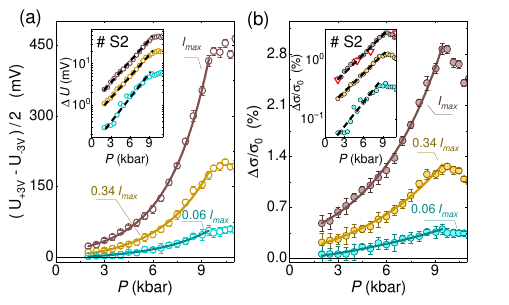}
	\caption{Pressure dependences of (a) the photoconductive signal $U = (U_{\rm +3V} -  U_{\rm -3V})/2$ and (b) the normalized photoconductivity $\Delta \sigma / \sigma_0$ for sample \#S2. The data were obtained for different values of the radiation intensity, indicated by the color code:  $I_{\rm max}=3.2$~kW/cm$^2$ (brown), $0.34\,I_{\rm max}$ (yellow), and $0.06\,I_{\rm max}$ (cyan). The insets show the results on a logarithmic scale. In addition to the data of the main panel we present the data from Fig.~\ref{Fig:4s1}(b) shown by down red triangles obtained for $I=3$~kW/cm$^2$. The solid lines show data fits with an exponential function.}
	\label{fig_PC_vs_P_normalizedS2}
\end{figure}

\begin{figure}[t]
	\includegraphics [width=\columnwidth, keepaspectratio] {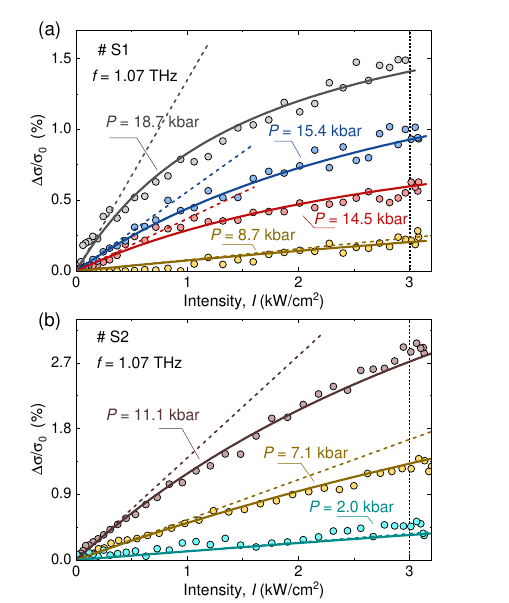} 	
	\caption{\label{Fig:4s1} 
		Intensity dependences of the normalized photoconductivity $\Delta\sigma/\sigma_0$ measured in  samples \#S1 (a) and \#S2 (b) under hydrostatic pressure ranging from ambient pressure to 18.7~kbar (a) and 11.1~kbar (b).  Solid lines are fits after Eqs.~\eqref{KDr} and~\eqref{DSSNL},  using $\Delta\sigma/\sigma_0 \propto K(I)\times I$. Dashed lines show  linear fits matching the low-intensity part of Eq.~\eqref{DSSNL}, $\Delta\sigma/\sigma_0 = A_{\rm 0} \times I$. The vertical dashed lines indicate the intensity of 3~kW/cm$^2$, at which the red down triangles shown in insets in Figs.~\ref{fig_PC_vs_P_normalizedS1}(b) and \ref{fig_PC_vs_P_normalizedS2}(b) are obtained.}
\end{figure}

\subsection{Methods and the radiation source}

The scheme of the experimental setup is shown in Fig.~\ref{Fig:1}(b). The photoconductivity was excited by normally incident monochromatic linearly polarized THz radiation. To measure the photoconductivity, an external DC bias $V= \pm 3$~V was applied to the circuit consisting of the sample and the load resistor $R_L = 1$~kOhm. The signal was picked up from the sample with the low resistance ($R << R_L$) and fed to a 50~MHz amplifier with a gain coefficient of 100.  As a radiation source we used a line-tunable  optically pumped pulsed molecular laser~\cite{Ganichev1982,Ganichev1995,Ganichev1998,Ganichev2005}.  The laser with NH$_3$ gas serving as the active media provided a linearly polarized radiation with $f = 1.07$~THz ($\lambda = 280$~$\mu$m, $\hbar \omega =4.4$~meV).  It operated in a pulsed regime with the pulse duration about 100~ns and the repetition rate of 1~Hz. The laser beam was focused using off-axis parabolic mirrors and controlled by a pyroelectric camera~\cite{Ganichev2002a}. It had an almost Gaussian shape with the spot diameter at the sample position about 1.5~mm. The peak power was measured by the fast photon-drag detector~\cite{Ganichev2006,Ganichev1985}.  The highest peak intensities of the THz pulses were about 200~kW/cm$^2$.  In experiments with samples placed in the pressure cell the radiation intensity was 40 times lower because of absorption in the pressure cell liquid and the small output aperture of the entrance cone, see Fig.~\ref{Fig:1pressure}. To vary the intensity of the laser radiation, we used two  wire grating polarizers: a first one was rotated, while the second one had a fixed position. This setup allowed us to vary the intensity of the radiation while ensuring that the output polarization remains unchanged~\cite{Hubmann2019,Candussio2021a}.

\begin{figure*}[t]
	\includegraphics [width=1.95\columnwidth, keepaspectratio] {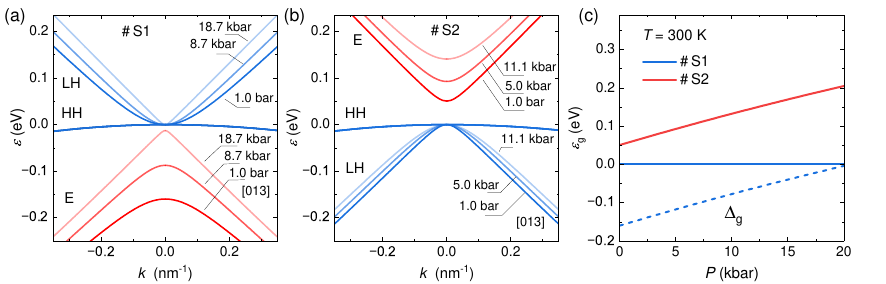}
	\caption{(a) and (b) Calculated band structures for samples \#S1 and \#S2 at room temperature under different hydrostatic pressures $P$ as marked. The red lines represent the subbands E derived from the $\Gamma_6$ state, while blue lines -- the  light hole (LH) and heavy hole (HH) subbands derived from the $\Gamma_8$ states. (c) The band gap energy $\varepsilon_g$ as a function of $P$ in samples \#S1 and \#S2 at room temperature (solid lines).  Note that sample \#S1 (HgTe) has $\varepsilon_g=0$ at all $P$, see (a). The blue dashed line shows the energy gap $\Delta_{g}$ between $\Gamma_6$ and $\Gamma_8$ bands at the $\Gamma$ point of the Brillouin zone in HgTe.}
	\label{Fig_k-p_calculations}
\end{figure*}

\section{\label{sec:results}Experimental results}

First, we present the data obtained at the atmospheric pressure. Figure~\ref{Fig:2}(a) shows the intensity dependence of the photoconductive signal caused by the THz radiation. The signal was obtained by measuring the response to positive and negative bias and calculating the photoconductive signal as $U = (U_{\rm +3V} -  U_{\rm -3V})/2$. In doing so, we eliminated a contribution of the photogalvanic current previously studied in Ref.~\cite{Hubmann2020}. Indeed, while the photocurrent is independent of the polarity of the bias voltage, the photoconductive response, according to Ohm's law, changes its sign when the polarity of the bias voltage is reversed. The data in Fig.~\ref{Fig:2}(a) show that at atmospheric pressure the signal in the sample with $x=0$ is nearly linear with the radiation intensity, while in the sample \#S2 ($x=0.13$) it initially grows linearly with increasing $I$ but the rate of increase in the photoconductive signal slows down as $I$ increases further (see dashed lines showing fitting data with linear intensity dependence). Figure~\ref{Fig:2}(b)  shows the relative change of the sample conductivity $\Delta \sigma = \sigma_i - \sigma_0$ normalized to the dark conductivity $\sigma_0$, where $\sigma_i$ is the conductivity under illumination. The relative change of the conductivity was calculated considering that $R\ll R_L$ after 
\begin{equation}
	\Delta \sigma / \sigma_0 \approx \frac{(U\cdot R_L)/(V\cdot R)}{1+(U\cdot R_L)/(V\cdot R)}.
\end{equation}
 Figure~\ref{Fig:2}(a) demonstrates that in both samples, THz radiation leads to an increase in conductivity, i.e., a decrease in sample resistance. 

Application of the hydrostatic pressure $P$ resulted in a significant increase in the magnitude of the signal and, consequently,  photoconductivity  $\Delta \sigma / \sigma_0$. The corresponding $P$-dependences are shown in Figs.~\ref{fig_PC_vs_P_normalizedS1} and ~\ref{fig_PC_vs_P_normalizedS2} for samples \#S1 ($x=0$) an \#S2 ($x=0.13$), respectively. Because of the observed nonlinearity, the data are presented for different radiation intensities. The data at the lowest value of $I$ correspond to the signals that are proportional to $I$, while the data at high intensities already correspond to the nonlinear regime. Solid lines in Figs.~\ref{fig_PC_vs_P_normalizedS1} and \ref{fig_PC_vs_P_normalizedS2}, as well as the dashed lines in the insets presenting the data in a log-lin plot, show that for pressure values above 3 kbar for sample \#S2 and 8 kbar for sample \#S1 the signal and $\Delta \sigma/ \sigma_0$ grow exponentially with $P$~\footnote{While the linear plot may give the impression that the two curves corresponding to reduced radiation intensity show a zero signal at pressures below 8~kbar, this is not actually the case. In reality, the signal is simply too weak to be distinguished from the noise. However, the same data plotted on a logarithmic scale are well described by three parallel straight lines over the entire range of available data, indicating no evidence of threshold behavior. Such behavior would also contradict our proposed theoretical model and numerical calculations.}. We note, that based on the current data, we cannot draw a definitive conclusion regarding the presence or absence of a threshold behavior. Clarifying this point would require further investigation, which is beyond the scope of the current work.
The apparent difference in the pressure dependences of the photoconductivity signal between samples \#S1 and \#S2 in the linear scale arises from the different absolute magnitudes of the measured signals and, consequently, from different relative signal-to-noise ratio. By contrast, when plotted on a logarithmic scale, the pressure dependence of photoconductivity for sample \#S2 is also well described by straight lines, with slopes close to those observed for sample \#S1.
While the data can be well fitted by an exponential growth for a wide pressure range, at very high $P$ and $I$ the behavior changes for sample \#S2: instead of continuing to grow, the $P$-dependence of the photoconductivity becomes weaker, see Fig.~\ref{fig_PC_vs_P_normalizedS2}.  In addition to enhancing the linear photoconductivity, application of the hydrostatic pressure significantly reduces the values of intensity at which the $I$-dependence becomes nonlinear, see Fig.~\ref{Fig:4s1} and Appendix~\ref{App:Anew}. In particular, at high $P$, the sublinear intensity dependence is clearly seen even in the sample \#S1 with $x=0$, see Fig.~\ref{Fig:4s1}(a). 

\section{Discussion}
\label{discussion}

First we discuss possible absorption mechanisms responsible for the photoconductive response. Figure~\ref{Fig_k-p_calculations} shows the band structure  calculated using the eight-band $\bm k\cdot \bm p$ Kane Hamiltonian, which directly takes into account the interactions between $\Gamma_6$, $\Gamma_8$, and $\Gamma_7$ bands (neglecting the contribution due to the bulk inversion asymmetry). Details of the explicit form of the Hamiltonian and pressure dependent band-structure parameters are provided in Ref.~\cite{Krishtopenko2016}. Figure~\ref{Fig_k-p_calculations}(c) demonstrates that in sample \#S2  the energy gap $\varepsilon_g$ exceeds the used photon energy even at atmospheric pressure and increases significantly when the hydrostatic pressure is applied. Thus, the absorption is possible solely due to indirect Drude-like optical transitions. Drude absorption is also dominates in sample \#S1 characterized by the zero energy gap. This is not surprising because at room temperature, the Fermi energy is several times higher than the photon energies used in our experiments, see below, and therefore, the interband absorption is negligible.

The Drude absorption of the radiation results in the heating of the electron gas, which changes the conductivity of the sample and leads to the photoconductive response. At room temperature, the average electron energy varies from 50-100~meV for sample \#S1 and 20 to 70~meV for sample \#S2. This means that the electron gas is non-degenerate and the electron-electron collision time is much shorter than the energy relaxation time caused by the electron-phonon interaction. Consequently, due to electron-electron scattering, the electron subsystem establishes an equilibrium energy distribution with a temperature $T_e$ that differs from the lattice temperature $T_0$. Under terahertz excitation and conditions above, the electron temperature can be found from the balance equation 
\begin{equation}
K(\omega)  I= N k_{\rm B}{T_e-T_0 \over \tau_\varepsilon},
\end{equation}
where $N$ is the free carrier concentration,  ${K(\omega)=N \mathcal S(\omega)}$ is the absorption coefficient, and $\mathcal S(\omega)$ the absorption cross section, which for $\omega\tau_p \gg 1$ 
is given by
\begin{equation}
	\label{K_0_T}
	\mathcal S(\omega) = {8\pi e^2 \over N mc n_\omega \omega^2 \tau_p} \sum_{\bm k}\qty[f_0(\varepsilon_k)-f_0(\varepsilon_k+\hbar\omega)].
\end{equation}
Here 
$\varepsilon_k$ is the energy of electrons with wavevector $\bm k$, $f_0$ is the equilibrium distribution function, $m$ is effective mass, $n_\omega$ refractive index, and $c$ the speed of light. Consequently we obtain
\begin{eqnarray}\label{e65}
\Delta T=T_e-T_0 \propto  {\tau_\varepsilon \over m \tau_p}.
\end{eqnarray}

The sample's conductivity is given by $\sigma =e N \mu$. The increase of the electron temperature results in the change of the electron mobility by $\Delta \mu = \mu(T_e) - \mu(T_0)$, and consequently, in the change of conductivity by $\Delta \sigma_\mu$ which, in the case of weak heating, can be well approximated by the simple expression
\begin{eqnarray}\label{e66}
	\frac{\Delta\sigma_\mu}{\sigma_0}= \frac{1}{\mu}\frac{\partial
		\mu}{\partial T_e}\bigg\vert_{T_e=T_0}\Delta T.
\end{eqnarray}
This mechanism is known as $\mu$-photoconductivity or electron bolometer, see, e.g., Ref.~\cite{Ganichev2005}.

In semiconductors with wide energy gaps, such as Ge, Si, or GaAs, and at high temperatures where all impurities are ionized, $\mu$-photoconductivity is the only mechanism responsible for the change in conductivity of the samples induced by electron gas heating. 
However, in our Cd$_x$Hg$_{1-x}$Te films with very narrow gap, another mechanism of photoconductivity may become possible. It is caused by an increase of the electron concentration with electron temperature, which is described by $\Delta N = N(T_e) - N(T_0)$ and, consequently: 
\begin{eqnarray}\label{e67}
	\frac{\Delta\sigma_N}{\sigma_0}= \frac{1}{N}\frac{\partial
		N}{\partial T_e}\bigg\vert_{T_e=T_0}\Delta T.
\end{eqnarray}
By analogy, we will refer to this mechanism as $N$-photoconductivity.

\begin{figure}[t]
	\centering \includegraphics[width=\linewidth]{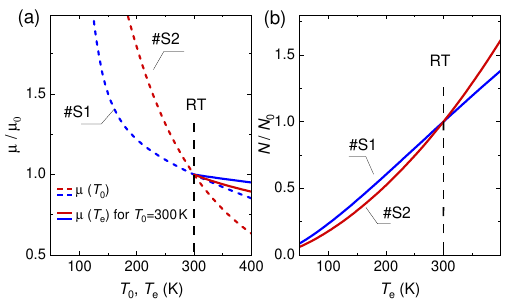} 	
	\caption{	$\mu-$ and $N-$mechanisms of photoconductivity.	(a) Calculated temperature dependence of electron mobility for samples \#S1 and \#S2, normalized to the mobility value $\mu_0$ at room temperature (RT). Scattering is assumed to be dominated by polar-optical phonon interaction, with the phonon energy of $\hbar \Omega = 19$~meV. The dashed lines represent the dependences under the assumption of equal electron and lattice temperatures, $T_e = T_0$. The curves obtained in this way show strong negative slope near RT. By contrast, if we reproduce the conditions of electron heating, i.e., fix $T_0 = 300$~K and plot the dependence on $T_e$, the resulting solid curves exhibit almost zero slope. (b) Calculated dependences of electron density $N(T_e)$,  normalized to $N_0=N(300\,\rm{K})$. Comparing the slopes of solid lines in panels (a) and (b), we conclude that the $N$-mechanism of photoconductivity dominates.}
\label{Fig_mu_vs_T}
\end{figure}

Strikingly, in all samples and for all pressure values we detected positive photoconductivity, i.e., the resistance drop upon irradiation. This is in contrast to THz photoconductivity in other materials (e.g. Ge or Si), for which $\mu$-photoconductivity is negative-- the resistance increases upon irradiation. At the first glance, this fact  also contradicts the observed temperature dependence of resistivity, shown in Fig.~\ref{Fig:2res}(a), and excludes the $\mu$-photoconductivity. However, the data in Fig.~\ref{Fig:2res}(a) are obtained via variation of the lattice temperature, which leads to the change of the number of phonons. By contrast, in our experiments, applying short ($\approx 150$~ns) single pulses with 1~Hz repetition frequency, only the temperature of electron gas  increases, while the lattice temperature remains unchanged.
 
Let us analyze whether the measured signal could be caused by $\mu$-photoconductivity. The results of the calculated dependence of $\mu(T_0)$ and $\mu(T_e)$ for the case of scattering on optical phonons, taking into account the realistic electron properties for each sample, are presented in Fig.~\ref{Fig_mu_vs_T}(a). The details of the calculation are provided in the Appendix. The figure shows that $\mu(T_0)$ rapidly decreases with increasing temperature. However, when considering electron heating with a fixed lattice temperature, the slope decreases significantly, and the dependence $\mu(T_e)$ becomes almost constant for both samples. Notably, calculations (not shown) also predict the possibility of increasing $\mu(T_e)$ for systems with a parabolic dispersion law. However, for the studied samples, due to the linear dispersion of electrons, the value of $\partial \mu(T_e)/\partial T_e$ remains negative in most experimental conditions and is always close to zero. Based on the above analysis, we conclude that $\mu$-photoconductivity is a possible but not the dominant mechanism because it is negative, which contradicts with experimental findings.

Positive $N$-photoconductivity requires an increase in the electron density by heating of the electron gas. In wide-gap semiconductors, due to the need for doping, this mechanism cannot be effective because of the pinning of the Fermi level to impurity levels. On the contrary, the films under study are narrow-gap and undoped, allowing for realization of intrinsic conductivity: the densities of electrons and holes coincide and are highly sensitive to temperature changes due to interband activation mechanism. 

The activation mechanism proceeds as follows: electrons in the conduction band absorb radiation and gain excess energy. Subsequently, through electron-electron, electron-hole, and hole-hole collisions, a new carrier distribution is formed with a quasi-equilibrium distribution function, characterized by an elevated carrier temperature and a shifted Fermi level. The valence band serves as a reservoir for the generation of additional electron-hole pairs, with holes carrier density equal to this of electrons – what is the  necessary  condition  to maintain overall charge neutrality of the system. The calculated temperature dependence of the electron (hole) density is shown in Fig.~\ref{Fig_Density_vs_T_P}(a). Interaction with the lattice is presumably not essential in this process but acoustic phonon emission/absorption may accelerate the establishment of electron -holes density quasi-equilibrium. The contribution of holes to the total conductivity is insignificant due to their larger effective mass and proportionally lower mobility, so in what follows we will focus on electrons (keeping in mind the overall charge neutrality). The calculated dependence of carrier density on electron temperature is shown in Fig.~\ref{Fig_mu_vs_T}(b). The details of the calculations are provided in the Appendix. The presented figure clearly shows that an increase of electron temperature leads to a significant rise in electron density that overcomes a smaller negative changes of mobility. We conclude that $N$-photoconductivity caused by electron gas heating dominates and leads to positive photoconductivity.

\begin{figure}[t]
	\centering \includegraphics[width=\linewidth]{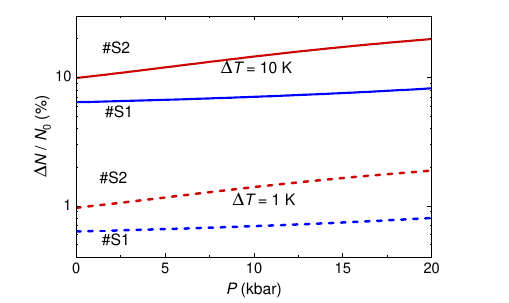} 	
	\caption{Calculated pressure dependences of the relative change of the electron density $\Delta N/N_0$ as a result of the electron heating by $\Delta T=1$~K (bottom traces) and 10~K (top traces). $N_0$ is electron density at the lattice temperature $T_0=300$~K. For absorption via the Drude mechanism, the absorbed power is proportional to the carrier concentration. Since $\Delta T$ reflects the absorbance normalized by the carrier concentration, it does not depend on $N_0$ and can be used as an input parameter in the calculation. At $\Delta T=1$~K, the photoconductivity amounts to 0.6-2~\%, which is consistent in order of magnitude with the experimental data and indicates that this estimate is realistic. However, the linear relationship between $\Delta T$ and $\Delta \sigma$ persists even for at least an order of magnitude stronger heating ($\Delta T = 10$~K).
    }
	\label{Fig_Density_and_Mobility_Change_vs_P}
\end{figure}

Let us analyze the behavior of photoconductivity as a function of pressure. Despite the measured exponential dependences of $\Delta \sigma / \sigma_0$ on $P$ (for more details see below) and its increase by more than an order of magnitude at the maximum applied pressure (see Fig.~\ref{fig_PC_vs_P_normalizedS1}), the calculation fails to explain this behavior. The computed dependences $\Delta N/N_0$ on $P$ in Fig.~\ref{Fig_Density_and_Mobility_Change_vs_P} show only a 20\% increase for sample \#S1. Sample \#S2 exhibits a more significant increase in $\Delta N/N_0$ by a factor of 2 at $P = 20$\,kbar. However, this is still not comparable to the experimental increase in photoconductivity by an order of magnitude, which also occurs at half the maximal pressure. 
We conclude that one needs another factor, which grows exponentially with $P$, and the only remaining possibility is an exponential increase in $\Delta T$ with pressure. As next, we show that the same conclusion can be drawn from the analysis of nonlinearities of the photoconductive response.

\begin{figure}[h]
	\includegraphics [width=\columnwidth, keepaspectratio]{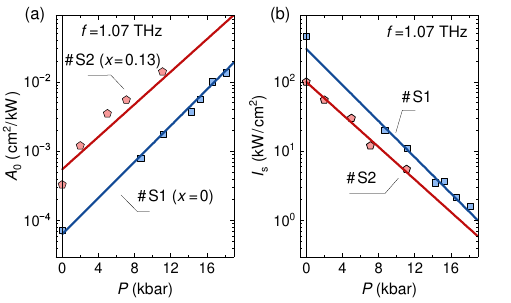}		\caption{\label{Fig:5S2}  Dependences of (a) the low intensity magnitude $A_0$ and (b) the saturation intensity $I_s$ on pressure $P$ obtained for samples \#S1 and \#S2 by fitting the data  in Figs.~\ref{Fig:2},\ref{Fig:4s1}, and \ref{Figadd} with Eq.~\eqref{DSSNL}. Solid lines are fits according to $A_0 \propto \exp(c / P)$ and $I_s \propto \exp(-c / P)$ yielding $c=0.3$ and 0.27 for samples \#S1 and \#S2, respectively.
	}
\end{figure}

Now we discuss the observed sublinear intensity dependence of the photoconductivity, see Figs.~\ref{Fig:2} and~\ref{Fig:4s1}.  We attribute it to electron gas heating resulting in bleaching of the Drude absorption and, consequently,  saturation of the electron temperature and photoconductive response. This process has been considered theoretically in Ref.~\cite{Moldavskaya2025}. It has been shown that  THz radiation induced  redistribution of carriers in the energy space results in the bleaching of the radiation absorption, which for $\omega\tau_p \gg 1$ is described by:
\begin{equation}
	\label{KDr}
	K(I)= {K_0\over 1+ I/I_s}, 
\end{equation}
with low intensity absorption coefficient  $K_0=N \mathcal S(\omega)$, $\mathcal S(\omega)$ given by by Eq.~\eqref{K_0_T}, and the saturation intensity
\begin{equation}
	\label{Is_Drude}
	I_s = {N k_{\rm B} \over \tau_\varepsilon \abs{\partial K_0/\partial T}} \propto {1\over \mathcal S(\omega)\tau_\varepsilon} \propto {m \tau_p \over \tau_\varepsilon}.
\end{equation}
This yields the intensity dependence of the relative  photoconductivity $\Delta \sigma /\sigma_0$:
\begin{equation}
	\label{DSSNL}
	{\Delta \sigma \over \sigma_0}  = {A_0 I \over 1+ I/I_s}, 
\end{equation}
where $A_0I$ is the low intensity magnitude of the relative photoconductivity.

Fitting of the data after Eq.~\eqref{DSSNL} shows that it describes well all experimental intensity dependences, see Figs.~\ref{Fig:2}(b) and \ref{Fig:4s1}. Dependences of the fitting parameters, the  amplitude $A_0$ and the saturation intensity $I_s$, on the pressure $P$ are shown in Fig.~\ref{Fig:5S2}. Strikingly, both values $A_0$ and $I_s$ depend exponentially on the pressure $P$~
\footnote{The observed reduction of the $\Delta \sigma / \sigma_0$ signal at high hydrostatic pressure $P$ is caused by the strong dependence of the saturation intensity $I_s$ on $P$, see Figs.~\ref{Fig:4s1} and \ref{Fig:5S2}(b). In the studied pressure range the saturation intensity drops by about two orders of magnitude with the pressure increase from the ambient one to 18~kbar (sample \#S1), and more than one order of magnitude by changing pressure from the ambient one to 12~kbar (sample \#S2), see Fig.~\ref{Fig:5S2}. The curves in the insets in Figs.~\ref{fig_PC_vs_P_normalizedS1} and \ref{fig_PC_vs_P_normalizedS2} are obtained for fixed radiation intensities. At high pressure they are substantially higher than $I_s$, therefore, the $\Delta \sigma / \sigma_0$ is measured already in the saturated regime. Consequently, this value is lower than expected for non-saturated regime and in the pressure dependencies they look as a decrease of the value from the exponential behavior. To obtain a correct pressure dependence of the $\Delta \sigma / \sigma_0$ we plotted in Fig.~\ref{Fig:5S2}(a) the values corresponding to the non-saturated magnitudes of the signal $A_0$ obtained from the fitting curves (see added dashed curves in Fig.~\ref{Fig:4s1} presenting $\Delta \sigma / \sigma_0 \propto A_0 \times I$). Figure~\ref{Fig:5S2}(a) shows that this value can be well fitted by the exponential function in the whole pressure range.}.  Furthermore, the exponents were found to have the same magnitude but opposite signs for both dependences.  This fact shows that the observed exponential dependences  have a  common origin:  they both are caused by the change of the electron gas heating.  Note that proportionality to the change of electron temperature of the $\Delta \sigma$ is characteristic for both $\mu$- and $N$-mechanisms of the photoconductivity, though we consider the $N$-photoconductivity as dominant one, see Eqs.~\eqref{e66} and \eqref{e67}.   We also obtained that the saturation intensity $I_s$ is proportional to $m\tau_p/\tau_\varepsilon$, from which it follows that  $I_s\propto 1/ \Delta T$, see  Eqs.~\eqref{e65}  and \eqref{Is_Drude}. Figure~\ref{Fig_k-p_calculations} shows that in the studied range of applied pressure $P$ the electron mass does not change significantly, thus the observed pressure dependence is almost exclusively caused by the exponential growth of the ratio of the energy and momentum relaxation times $\tau_\varepsilon /\tau_p$. Furthermore, since the electron mobility changes only weakly with the pressure increase, our data show that the hydrostatic pressure strongly affects the energy relaxation time, which controls both the electron temperature and the saturation intensity. The microscopic  origin of this surprising observation requires further studies.

\section{Summary} 
\label{summary}

In summary, we demonstrate that in Cd$_x$Hg$_{1-x}$Te films the absorption of THz radiation leads to positive photoconductivity at room temperature. The photoconductivity mechanism is associated with Drude absorption of radiation, which leads to electron gas heating and a subsequent rise in carrier concentration. Such an increase is ascribed to an interband activation mechanism, specific for narrow-gap undoped semiconductors. The observed photoresponse saturates as the radiation intensity increases. It is shown that the saturation is due to  the absorption bleaching caused by the electron gas heating. Strikingly under hydrostatic pressure $P$ the magnitude of the photoconductivity increases exponentially with $P$ and the saturation intensity exponentially decreases with $P$. Importantly, both dependences are described by the same exponent, with only a slight difference between samples with $x=0$ and $x=0.13$. We show that the magnitude of photoconductivity and saturation intensities are controlled by the increase in electron temperature $\Delta T \propto \tau_\varepsilon / \tau_p$. Further studies are needed to understand the microscopic origin of this result.

\section{Acknowledgments}
\label{acknow}
The authors thank N.~N.~Mikhailov and S.~A.~Dvoretsky for providing high quality Cd$_x$Hg$_{1-x}$Te material and valuable discussions.  We acknowledge the financial support of the Deutsche Forschungsgemeinschaft (DFG, German Research Foundation) via  Project-ID 521083032 (Ga501/19), of the French Agence Nationale pour la Recherche for TEASER project (ANR-24-CE24-4830), and the France 2030 program through and Equipex+ HYBAT project (ANR-21-ESRE-0026). S.D.G. and I.Y. are grateful for the support of the European Union (ERC-ADVANCED TERAPLASM \#101053716). I.A.D. acknowledges support of the DFG via project DM1/6-1. We also acknowledge the support of “Center for Terahertz Research and Applications (CENTERA2) project (FENG.02.01-IP.05-T004/23).

\appendix
\counterwithin{figure}{section}
\setcounter{figure}{0}

\section{Additional data}
\label{App:Anew}

\begin{figure}[t]
	\includegraphics [width=\columnwidth, keepaspectratio] {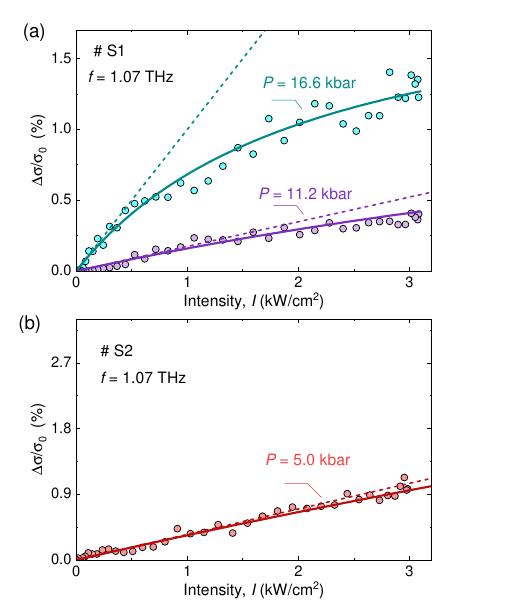} 	
	\caption{\label{Figadd} 
		Intensity dependences of the normalized photoconductivity $\Delta\sigma/\sigma_0$ measured in  samples \#S1, panel (a), and \#S2, panel (b), for several values of the  hydrostatic pressure.  Solid lines are fits after Eqs.~\eqref{KDr} and~\eqref{DSSNL},  using $\Delta\sigma/\sigma_0 \propto K(I)\times I$. Dashed lines show   linear fits matching the low-intensity part of Eq.~\eqref{DSSNL}, $\Delta\sigma/\sigma_0 = A_{\rm 0} \times I$.}
\end{figure}

	Figure~\ref{Figadd} shows intensity dependences of the normalized photoconductivity $\Delta\sigma/\sigma_0$ measured in  samples \#S1, panel (a), and \#S2, panel (b), for several values of the  hydrostatic pressure.

\section{Density calculations}

\label{App:A}

\begin{figure}[t]
	\centering \includegraphics[width=\linewidth]{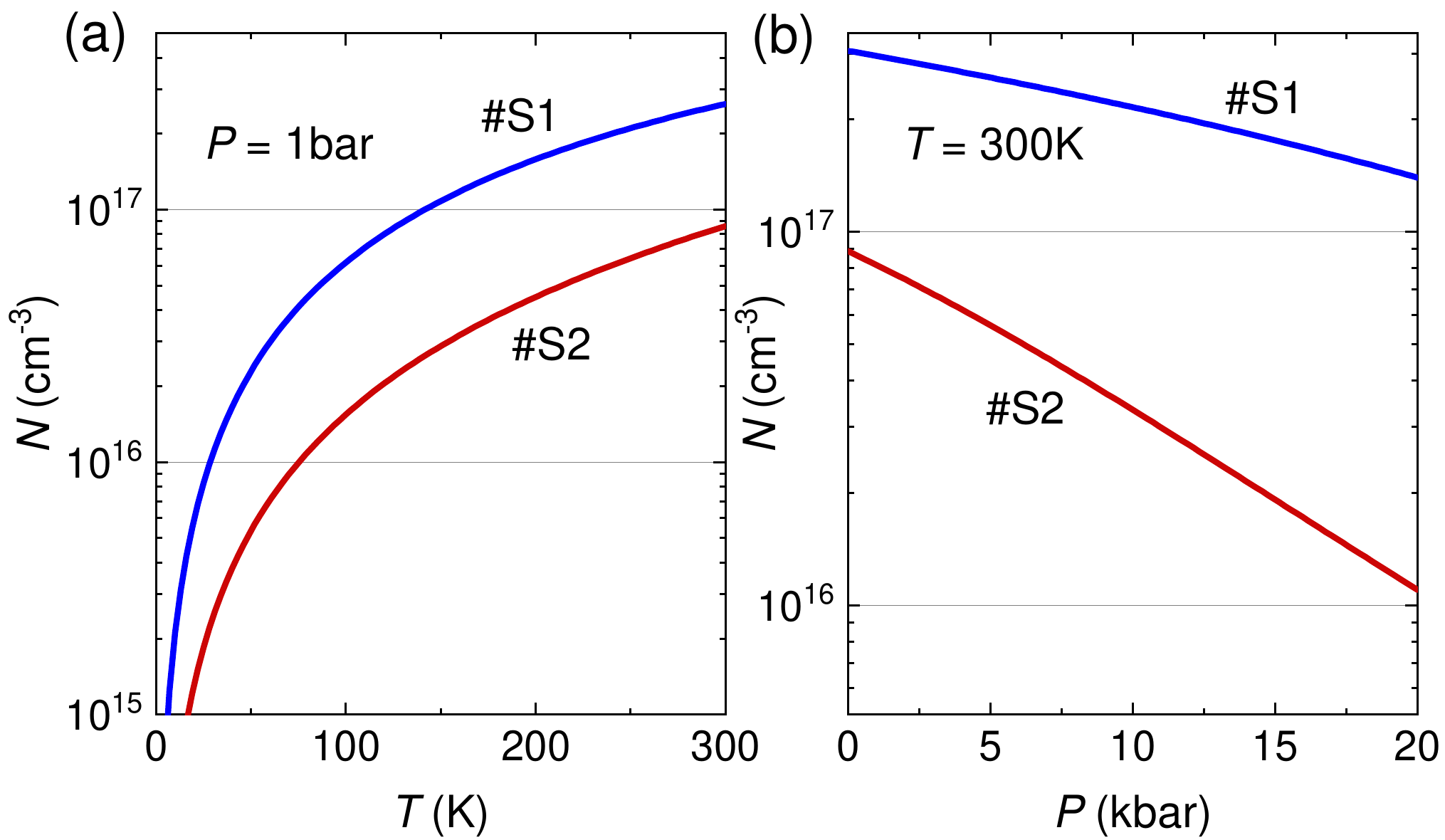} 	
	\caption{ Calculated electron density $N$ in the studied samples. (a) Temperature dependence at ambient pressure. (b) Dependence on hydrostatic pressure at a fixed $T=300$\,K. Calculations were performed using the actual band structure from Fig.~\ref{Fig_k-p_calculations}, assuming the absence of band bending, impurities, and overall charge neutrality of the system implying an equal number of free electrons and holes determined by the spectrum and temperature.}
	\label{Fig_Density_vs_T_P}
\end{figure}

We start with the description of electron density calculations. The calculations take into account the actual band structure for the given Cd fraction $x$ in the film composition, the lattice temperature $T_0$, and the pressure $P$. For the calculation of the equilibrium (dark) density (see Fig.~\ref{Fig_Density_vs_T_P}) the electron temperature $T_e$ was taken equal to the lattice temperature $T_0$. However, when analyzing the effects of electron heating on density or mobility (see Figs.~\ref{Fig_mu_vs_T} and \ref{Fig_Density_and_Mobility_Change_vs_P}), $T_e$ was treated as a separate parameter and accounted for through the electron Fermi-Dirac  distribution function. 

The calculations were performed under the assumption of overall charge neutrality and within the intrinsic semiconductor approximation, assuming that the densities of electrons and holes are equal. This approximation holds as long as charged residual impurities do not significantly affect the electron density and the Fermi level position $E_F$. While it may break down at liquid helium temperatures, it definitely holds at room temperature when $k_\text{B} T_e \sim \varepsilon_g$. The presence of holes is essential for maintaining the charge neutrality; however, their contribution to (photo-)conductivity is negligible compared to that of electrons due to their larger effective mass and lower mobility. According to the calculations, light holes are also present in the system, but their contribution to conductivity remains insignificant due to their low density.

The obtained in this way dependences of the dark density $N$ on temperature $T_0$, as shown in Fig.~\ref{Fig_Density_vs_T_P}(a), are consistent with the experimental data for $R(T_0)$ presented in Fig.~\ref{Fig:2res}(a).  As temperature increases from 4~K to 300~K, the electron density increases by 2-3 orders of magnitude while the mobility decreases (Fig.~\ref{Fig_mu_vs_T}). As a result of their partial compensation, a relatively weak and nonmonotonic temperature dependence of $R(T)$ is observed. The calculation of electron density as a function of pressure agrees with the experiment not only qualitatively but also quantitatively: according to Fig.~\ref{Fig:2res}(b), an increase in pressure leads to a twofold increase in resistance for sample \#S1 and an order of magnitude increase for sample \#S2;  the density decreases for both samples, see Fig.~\ref{Fig_Density_vs_T_P}(b). At the same time, only a weak dependence of mobility (less then several tens of percent, not shown) vs $P$ is predicted by calculations, making it a negligible factor.

\begin{figure}[t]
	\centering \includegraphics[width=\linewidth]{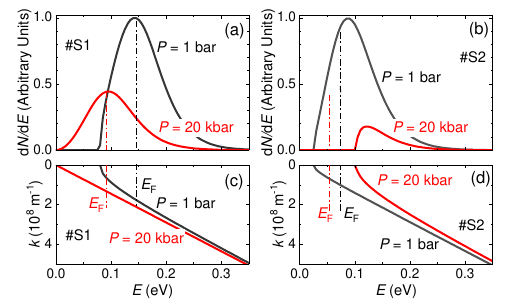} 	
	\caption{ Calculated distribution of electron density over energies $\text{d}N/\text{d}E$ vs $E$ for samples \#S1 (a) and \#S2 (b) for ambient pressure (black) and $P = 20$~kbar (red) at $T = 300$~K. (c) and (d) Corresponding dispersion relations of the conduction band. The vertical lines indicate the position of the Fermi level.}
	\label{Fig_Density_Distribution}
\end{figure}

On the other hand, pressure significantly affects the electronic bands. In Fig.~\ref{Fig_Density_Distribution}, top panels (a) and (b) show the calculated distribution of electron density over energies for samples \#S1 and \#S2 at ambient pressure and at $P = 20$~kbar. The bottom panels display the corresponding conduction band dispersion on the same energy scale, along with the position of the Fermi level. A key difference between samples \#S1 and \#S2 is that, at all pressures, the Fermi level in \#S1 remains in the conduction band, positioned at between 2 and 3 $k_\text{B} T$ above its bottom. By contrast, in \#S2, as pressure increases, the Fermi level shifts into the bandgap, leading to a sharp decrease in electron density and a Boltzmann-like energy distribution. A second important distinction is that for \#S1 at $P = 20$~kbar, the spectrum becomes nearly linear, whereas for \#S2, it remains parabolic. Importantly, strong modifications of the electronic spectrum with $P$ do not lead to significant changes in the $\Delta N/ N_0$ (see Fig.~\ref{Fig_Density_and_Mobility_Change_vs_P}) and could not explain the observed experimental data without introducing of exponential dependence of $\Delta T$ vs $P$.

\section{Mobility calculations}

The calculations of electron mobility under dominant mechanism of scattering by optical phonons were performed numerically, depending on the lattice $T_0$ and electron $T_e$ temperatures and pressure $P$, taking into account the real spectrum and the position of the Fermi level at the given $T_e$ for each sample. According to Ref.~\cite{Jena2022}, the total momentum scattering rate on polar optical phonons for electron with energy $\varepsilon_k$ is given by the formula:
\begin{eqnarray}\label{scattering_rate}
	1/\tau_m = \Omega \alpha_{LO} N_{LO} \times \nonumber \\
	\sqrt{x} 
	\Big[
	\left(\sqrt{1+x}-x \sinh^{-1}x^{-1/2}\right) + \nonumber \\
	e^z \left(\sqrt{1-x} + x \sinh^{-1}\sqrt{x^{-1}-1} \right)\Theta(1-x) \Big]
\end{eqnarray}
%
%
where $\tau_m$ is electron momentum relaxation time, $\alpha_{LO}$ is electron-phonon coupling coefficient, $N_{LO}=1/(e^z-1)$ with $z = \hbar \Omega / (k_\text{B} T_0)$ is the optical phonon occupation function, $x = \hbar \Omega / \varepsilon_k$ is a relation between optical phonon energy and electron kinetic energy $\varepsilon_k$, $\Theta(x)$ is the Heaviside step function and $\sinh^{-1}(x)$ is an inverse hyperbolic sine function. 

In order to proceed to the mobility, one needs the electron energy dispersion $\varepsilon_k = \sqrt{(\hbar k c_{DF})^2 + (\varepsilon_g/2)^2}-\varepsilon_g/2$, where $k$ is an electron momentum, $c_{DF} = 0.0035 \times c$ is a numerical constant (speed of Dirac fermions at large energies, $c$ is a speed of light in vacuum), and $\varepsilon_g(P,T_0,x)$ is an energy gap between the LH and E bands. Next, we introduce the electron density of states $\nu (\varepsilon_k) = (\varepsilon_k+\varepsilon_g/2)\sqrt{ \varepsilon_k (\varepsilon_g+\varepsilon_k)}/(c_{DF}^3 \hbar^3 \pi^2)$, group velocity  $v = c_{DF} \sqrt{\varepsilon_k (\varepsilon_k + \varepsilon_g)}/(\varepsilon_k + \varepsilon_g/2)$ and electron Fermi-Dirac distribution function $f_0(\varepsilon_k,E_F,T_e)$. The electron temperature is taken into account in two ways: directly through the value of $T_e$ in the distribution function and through the dependence of the Fermi level $E_F(T_e)$, whose position is determined by the charge neutrality condition. 

Finally, the mobility can be obtained using the formula:
\begin{eqnarray}\label{mobility}
	\mu \propto \frac{\int_{0}^{\infty} v(\varepsilon_k)^2 \nu(\varepsilon_k) \left(-\frac{df_0(\varepsilon_k,E_F,T_e)}{d\varepsilon_k}\right) \tau_m(\varepsilon_k) d\varepsilon_k}
	{\int_{0}^{\infty} \nu(\varepsilon_k) f_0(\varepsilon_k,E_F,T_e) d\varepsilon_k },
	 \nonumber
\end{eqnarray}
where we have omitted numerical coefficients since we were only interested in functional dependences.

\bibliography{all_lib1.bib}
\end{document}